# Statistics of narrow-band partially polarized light

## MIKHAIL CHARNOTSKII[*]

 *mikhail.charnotskii@gmail.com*

**A complete single-point statistical description of a narrow-band partially polarized optical field is developed in terms of the 2-D Period-Averaged Probability Density Function (PA-PDF) of the electrical field vector. This statistic can be measured using the coherent (heterodyne) detection. PA-PDF carries more information about the partially-polarized light than the traditional Stokes vector. For a simple Gaussian partially polarized field the PA-PDF depends on 13 real parameters in contrast to the four parameters of the Stokes vector or coherence tensor.**

## 1. Introduction

For more than 150 years, the Stokes vector [1] has been exclusively used to describe the local polarization state of electromagnetic radiation. The most recent review paper on the subject [2] still states that "*polarization … is fully described by the … Stokes vector.*" The popularity of the Stokes parameters is based on the fact that a linear polarizer and retardation plate are all that is needed to measure them [3]. The more recent development of the so-called unified theory of coherence and polarization of stochastic electromagnetic beams, see for example [4], added spatial coherence to the polarization theory and made it possible to investigate the propagation of partially-polarized beam waves. However at each point the coherence tensor of the field, which is called a cross-spectral density matrix in [4], contains the same information as the Stokes vector.

It is clear from the beginning that the Stokes vector, which is comprised only from the second statistical moments of the field, cannot provide exhaustive information about the statistics of the polarized field. Even when only the first and the second statistical moments of the field are considered, there are as many as 13 parameters that characterize the statistics.

In Section 2 we present statistics of the narrow band partially-polarized field in terms of the nonstationary oscillating Probability Density Function (PDF), introduce the more perceivable concept of the Period-Averaged PDF (PA-PDF), and discuss the relationship between the classic Stokes vector and the statistical moments of our description. In Section 3, we apply our approach to the simplest case of the oscillating bivariate Gaussian PDF and show some examples of the ensuing PA-PDFs. In Section 4, based on several numerical examples, we show that the Stokes vector does not fully characterize the state of the partially polarized wave. The discussion in Section 5 primarily addresses the measurements of the PA-PDF and associated statistical moments.

## 2. General Statistics

Electric field $\mathbf{E}(t)$ of the transverse electromagnetic wave with carrier frequency $\omega$ propagating along the *z*-axis at some point in the $(x,y)$ plane can be presented as

$$\mathbf{E}(t) = \begin{pmatrix} E_x(t) \\ E_y(t) \end{pmatrix} = \mathrm{Re}\begin{pmatrix} u(t)e^{i\omega t} \\ v(t)e^{i\omega t} \end{pmatrix}. \qquad (1)$$

Here $E_x$ and $E_y$ are orthogonal components of an electrical field in the *x* and *y* directions, and $u(t)$ and $v(t)$ are random complex amplitudes of these components. For the quasi-monochromatic waves considered here, correlation time $t_C$ of $u(t)$ and $v(t)$ is much larger than the carrier oscillation period

$$\omega t_C \gg 1. \qquad (2)$$

Field $\mathbf{E}(t)$ is a real-valued two-dimensional nonstationary random vector that can be represented as

$$\mathbf{E}(t) = \begin{pmatrix} u_R(t)\cos(\omega t) - u_I(t)\sin(\omega t) \\ v_R(t)\cos(\omega t) - v_I(t)\sin(\omega t) \end{pmatrix}. \qquad (3)$$

where subscripts *R* and *I* stand for the real and imaginary parts (in-phase and quadrature components) of corresponding complex amplitudes. For time intervals less than $t_C$ vector $\mathbf{E}(t)$ inscribes an ellipse at the $(E_x, E_y)$ plane. For a narrow-band polarized wave this ellipse randomly changes with time. These changes are slow in the carrier oscillations time scale $t_O = 2\pi/\omega$. Figure 1 shows two examples of hodographs of vector $\mathbf{E}(t)$. Dark ellipses represent the mean or coherent component $\langle \mathbf{E}(t) \rangle$. Here and further on the angular brackets indicate the statistical averaging over the field fluctuations. For the ease of graphical presentation a relatively short correlation time $t_C/t_O = 3$ was chosen here, and 20 periods of carrier oscillations were traced. In the case of relatively small field fluctuations with respect to the mean shown at the left panel, $\mathbf{E}(t)$ remains close to the polarization ellipse of the mean field. The right panel's field fluctuations are larger than the mean field and $\mathbf{E}(t)$ outlines ellipses, which shape, size and orientation slowly evolve with time.

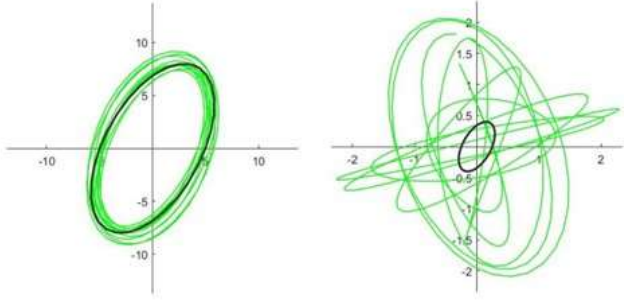

Fig. 1 Two examples of the electric vector hodographs for partially-polarized waves. Left panel: mean field is larger than the fluctuating component. Right panel: mean field is smaller than the fluctuating component.

The complete statistic of the random vector $\mathbf{E}(t)$ at any fixed moment is given by nonstationary Probability Density Function (PDF)

$$P(E_x, E_y, t)dE_x dE_y = Probab(E_x < E_x(t) < E_x + dE_x, E_y < E_y(t) < E_y + dE_y). \quad (4)$$

Obviously $P(E_x, E_y, t)$ is a $t_O$-periodic function of time. Typically, the carrier frequency is known, and carries no useful information. The classical statistics of a polarized field, such as Jones vector,

$$\mathbf{J} \equiv \begin{pmatrix} u \\ v \end{pmatrix} = \begin{pmatrix} u_R + i u_I \\ v_R + i v_I \end{pmatrix}, \quad (5)$$

Stokes vector

$$\mathbf{S} \equiv \begin{pmatrix} I \\ Q \\ U \\ V \end{pmatrix} = \begin{pmatrix} \langle |u|^2 \rangle + \langle |v|^2 \rangle \\ \langle |u|^2 \rangle - \langle |v|^2 \rangle \\ 2\operatorname{Re}\langle uv^* \rangle \\ -2\operatorname{Im}\langle uv^* \rangle \end{pmatrix} = \begin{pmatrix} \langle u_R^2 \rangle + \langle u_I^2 \rangle + \langle v_R^2 \rangle + \langle v_I^2 \rangle \\ \langle u_R^2 \rangle + \langle u_I^2 \rangle - \langle v_R^2 \rangle - \langle v_I^2 \rangle \\ 2\langle u_R v_R \rangle + 2\langle u_I v_I \rangle \\ 2\langle u_R v_I \rangle - 2\langle u_I v_R \rangle \end{pmatrix}, \quad (6)$$

and correlation tensor

$$\mathbf{W} \equiv \begin{pmatrix} \langle |u|^2 \rangle & \langle uv^* \rangle \\ \langle u^* v \rangle & \langle |v|^2 \rangle \end{pmatrix}$$

$$= \begin{pmatrix} \langle u_R^2 \rangle + \langle u_I^2 \rangle & \langle u_R v_R \rangle + \langle u_I v_I \rangle - i(\langle u_R v_I \rangle - \langle u_I v_R \rangle) \\ \langle u_R v_R \rangle + \langle u_I v_I \rangle + i(\langle u_R v_I \rangle - \langle u_I v_R \rangle) & \langle v_R^2 \rangle + \langle v_I^2 \rangle \end{pmatrix} \quad (7)$$

are all period-averaged statistics. The Stokes vector and correlation tensor hold the same information about the statistics of partially-polarized field, and the Stokes vector can be represented in terms of the components of the correlation tensor and vice versa.

Here we assume that it is possible to measure the instantaneous components of the field, but the exact time of the measurement is undetermined at the carrier oscillation scale $t_O$. In other words, the absolute phase of the field is unknown or irrelevant. Alternatively, the multiple measurements of $(E_x, E_y)$ are performed at the time moments $t_n$ that are uniformly distributed over the carrier oscillation period $t_O$. In this case the probability distribution of the measurement outcomes is the Period-Averaged PDF (PA-PDF)

$$\overline{P}(E_x, E_y) \equiv \frac{1}{t_O} \int_0^{t_O} dt P(E_x, E_y, t). \quad (8)$$

PA-PDF contains complete statistics of the $\mathbf{E}(t)$ except the part that is related to the absolute phase, or clock zero. For example, PA-PDF cannot provide probability distribution of the field at $t=1.33$ seconds. However, if the field is measured at arbitrary time moments, PA-PDF gives the probable distribution of the outcomes.

Given the joint probability distribution of the quadrature components $w(u_R, u_I, v_R, v_I)$ it is possible to calculate the non-stationary PDF of the field as

$$P(E_x, E_y, t) = \langle \delta(E_x - u_R \cos\omega t + u_I \sin\omega t) \delta(E_y - v_R \cos\omega t + v_I \sin\omega t) \rangle \quad (9)$$
$$= \iint d\xi d\eta w(E_x \cos\omega t + \xi \sin\omega t, -E_x \sin\omega t + \xi \cos\omega t, E_y \cos\omega t + \eta \sin\omega t, -E_y \sin\omega t + \eta \cos\omega t),$$

and then use Eq. (8) to calculate the PA_PDF.

Obviously, the Stokes vector does not exhaust the statistics of the field. For instance, the Stokes vector does not include the statistical moments of the orders higher than two. But even when only the moments up to the second order are concerned, the Stokes vector still does not capture all the information. The deficiency of the Stokes vector description was noticed before [5, p. 24], but no attempt was made to develop the comprehensive statistics.

Presenting the in-phase and quadrature parts of the complex amplitudes as sums of mean and the fluctuating components

$$u_R(t) = \langle u_R \rangle + \widetilde{u}_R(t), \; u_I(t) = \langle u_I \rangle + \widetilde{u}_I(t),$$
$$v_R(t) = \langle v_R \rangle + \widetilde{v}_R(t), \; v_I(t) = \langle v_I \rangle + \widetilde{v}_I(t), \quad (10)$$

we limit our attention to the mean values $\langle u_R \rangle$, $\langle u_I \rangle$, $\langle v_R \rangle$ and $\langle v_I \rangle$, and the covariance matrix

$$\mathbf{C} \equiv \begin{pmatrix} \langle \widetilde{u}_R^2 \rangle & \langle \widetilde{u}_R \widetilde{u}_I \rangle & \langle \widetilde{u}_R \widetilde{v}_R \rangle & \langle \widetilde{u}_R \widetilde{v}_I \rangle \\ \langle \widetilde{u}_R \widetilde{u}_I \rangle & \langle \widetilde{u}_I^2 \rangle & \langle \widetilde{u}_I \widetilde{v}_R \rangle & \langle \widetilde{u}_I \widetilde{v}_I \rangle \\ \langle \widetilde{u}_R \widetilde{v}_R \rangle & \langle \widetilde{u}_I \widetilde{v}_R \rangle & \langle \widetilde{v}_R^2 \rangle & \langle \widetilde{v}_R \widetilde{v}_I \rangle \\ \langle \widetilde{u}_R \widetilde{v}_I \rangle & \langle \widetilde{u}_I \widetilde{v}_I \rangle & \langle \widetilde{v}_R \widetilde{v}_I \rangle & \langle \widetilde{v}_I^2 \rangle \end{pmatrix}. \quad (11)$$

The Stokes vector can be represented in terms of the mean values and covariances as

$$\mathbf{S} = \begin{pmatrix} \langle u_R \rangle^2 + \langle u_I \rangle^2 + \langle v_R \rangle^2 + \langle v_I \rangle^2 + \langle \widetilde{u}_R^2 \rangle + \langle \widetilde{u}_I^2 \rangle + \langle \widetilde{v}_R^2 \rangle + \langle \widetilde{v}_I^2 \rangle \\ \langle u_R \rangle^2 + \langle u_I \rangle^2 - \langle v_R \rangle^2 - \langle v_I \rangle^2 + \langle \widetilde{u}_R^2 \rangle + \langle \widetilde{u}_I^2 \rangle - \langle \widetilde{v}_R^2 \rangle - \langle \widetilde{v}_I^2 \rangle \\ 2\langle u_R \rangle\langle v_R \rangle + 2\langle u_I \rangle\langle v_I \rangle + 2\langle \widetilde{u}_R \widetilde{v}_R \rangle + 2\langle \widetilde{u}_I \widetilde{v}_I \rangle \\ 2\langle u_R \rangle\langle v_I \rangle - 2\langle u_I \rangle\langle v_R \rangle + 2\langle \widetilde{u}_R \widetilde{v}_I \rangle - 2\langle \widetilde{u}_I \widetilde{v}_R \rangle \end{pmatrix}. \quad (12)$$

It is immediately noticeable that the covariances $\langle \widetilde{u}_R \widetilde{u}_I \rangle$ and $\langle \widetilde{v}_R \widetilde{v}_I \rangle$ do not affect the Stokes vector, and the parameters' count suggests that only four combinations of the possible 14 values of the four means and ten components of the symmetric Semi-Positive

Definite (SPD) covariance matrix $\mathbf{C}$ enter the Stokes vector. Note that the 14 components are essential for the oscillating non-stationary PDF, Eq. (4). Since the field state at $t=0$ is not relevant for the PA-PDF, Eq. (8), we can always eliminate one of the 14 parameters, e. g. by setting $\langle E_y(0)\rangle = 0$ or, equivalently, $v_R = 0$. This leaves 13 parameters that fully define the first and second-order statistical moments of the polarization state of the partially-polarized wave in contrast to the four parameters offered by the Stokes vector's description.

Correlation tensor $\mathbf{W}$, Eq. (7) can be presented in terms of the mean values and covariances as

$$\mathbf{W} = \begin{pmatrix} \langle u_R\rangle^2 + \langle u_I\rangle^2 \\ + \langle \tilde{u}_R^2\rangle + \langle \tilde{u}_I^2\rangle \end{pmatrix} \begin{pmatrix} \langle u_R\rangle\langle v_R\rangle - \langle u_I\rangle\langle v_I\rangle \\ + \langle \tilde{u}_R\tilde{v}_R\rangle - \langle \tilde{u}_I\tilde{v}_I\rangle \\ + i(\langle u_R\rangle\langle v_I\rangle + \langle u_I\rangle\langle v_R\rangle \\ + \langle \tilde{u}_R\tilde{v}_I\rangle + \langle \tilde{u}_I\tilde{v}_R\rangle) \end{pmatrix} \\ \begin{pmatrix} \langle u_R\rangle\langle v_R\rangle - \langle u_I\rangle\langle v_I\rangle \\ + \langle \tilde{u}_R\tilde{v}_R\rangle - \langle \tilde{u}_I\tilde{v}_I\rangle \\ - i(\langle u_R\rangle\langle v_I\rangle + \langle u_I\rangle\langle v_R\rangle \\ + \langle \tilde{u}_R\tilde{v}_I\rangle + \langle \tilde{u}_I\tilde{v}_R\rangle) \end{pmatrix} \begin{pmatrix} \langle v_R\rangle^2 + \langle v_I\rangle^2 \\ + \langle \tilde{v}_R^2\rangle + \langle \tilde{v}_I^2\rangle \end{pmatrix} . \quad (13)$$

It is customary to present the Stokes vector as a sum of polarized and non-polarized parts

$$\mathbf{S} = \mathbf{S}_{POL} + \mathbf{S}_{DEP}$$
$$= \begin{pmatrix} \sqrt{Q^2 + U^2 + V^2} \\ Q \\ U \\ V \end{pmatrix} + \begin{pmatrix} I - \sqrt{Q^2 + U^2 + V^2} \\ 0 \\ 0 \\ 0 \end{pmatrix}. \quad (14)$$

Eq. (10) suggests an alternative presentation of the Stokes vector as a sum of the mean and fluctuating components

$$\mathbf{S} = \mathbf{S}_M + \mathbf{S}_F$$
$$= \begin{pmatrix} \langle u_R\rangle^2 + \langle u_I\rangle^2 + \langle v_R\rangle^2 + \langle v_I\rangle^2 \\ \langle u_R\rangle^2 + \langle u_I\rangle^2 - \langle v_R\rangle^2 - \langle v_I\rangle^2 \\ 2\langle u_R\rangle\langle v_R\rangle + 2\langle u_I\rangle\langle v_I\rangle \\ 2\langle u_R\rangle\langle v_I\rangle - 2\langle u_I\rangle\langle v_R\rangle \end{pmatrix} \quad (15)$$
$$+ \begin{pmatrix} \langle \tilde{u}_R^2\rangle + \langle \tilde{u}_I^2\rangle + \langle \tilde{v}_R^2\rangle + \langle \tilde{v}_I^2\rangle \\ \langle \tilde{u}_R^2\rangle + \langle \tilde{u}_I^2\rangle - \langle \tilde{v}_R^2\rangle - \langle \tilde{v}_I^2\rangle \\ 2\langle \tilde{u}_R\tilde{v}_R\rangle + 2\langle \tilde{u}_I\tilde{v}_I\rangle \\ 2\langle \tilde{u}_R\tilde{v}_I\rangle - 2\langle \tilde{u}_I\tilde{v}_R\rangle \end{pmatrix}.$$

The mean component is always fully polarized. The fluctuating component may or may not have a polarized part.

### 3. PA-PDF for a Gaussian polarized field

In this section we illustrate the conceptual development of the previous section by, possibly, the simplest example of normal distribution of the complex amplitudes of the field. Namely, we assume that

$$w(u_R, u_I, v_R, v_I)$$
$$= \frac{1}{4\pi^2 \sqrt{|\mathbf{C}|}} \exp\left[-\frac{1}{2}\begin{pmatrix} u_R - \langle u_R\rangle \\ u_I - \langle u_I\rangle \\ v_R - \langle v_R\rangle \\ v_I - \langle v_I\rangle \end{pmatrix}^T \mathbf{C}^{-1} \begin{pmatrix} u_R - \langle u_R\rangle \\ u_I - \langle u_I\rangle \\ v_R - \langle v_R\rangle \\ v_I - \langle v_I\rangle \end{pmatrix}\right]. \quad (16)$$

Nonstationary PDF $P(E_x, E_y, t)$, Eq. 4, can be calculated using Eq. (8). However, since, according to Eq. (3), the field components $E_x$ and $E_y$ are linear combinations of the normal complex amplitudes, they also have normal distribution. Parameters of this normal distribution can be calculated as follows. The mean field is

$$\langle \mathbf{E}(t)\rangle = \begin{pmatrix} \langle E_x(t)\rangle \\ \langle E_y(t)\rangle \end{pmatrix} = \begin{pmatrix} \langle u_R\rangle C - \langle u_I\rangle S \\ \langle v_R\rangle C - \langle v_I\rangle S \end{pmatrix}, \quad (17)$$

and the covariance tensor is

$$\begin{pmatrix} \langle \tilde{E}_x^2(t)\rangle & \langle \tilde{E}_x(t)\tilde{E}_y(t)\rangle \\ \langle \tilde{E}_x(t)\tilde{E}_y(t)\rangle & \langle \tilde{E}_y^2(t)\rangle \end{pmatrix}$$
$$= \begin{pmatrix} \langle \tilde{u}_R^2\rangle C^2 + \langle \tilde{u}_I^2\rangle S^2 & \langle \tilde{u}_R\tilde{v}_R\rangle C^2 + \langle \tilde{u}_I\tilde{v}_I\rangle S^2 \\ -2\langle \tilde{u}_R\tilde{u}_I\rangle CS & -[\langle \tilde{u}_R\tilde{v}_I\rangle + \langle \tilde{u}_I\tilde{v}_R\rangle]CS \\ \langle \tilde{u}_R\tilde{v}_R\rangle C^2 + \langle \tilde{u}_I\tilde{v}_I\rangle S^2 & \langle \tilde{v}_R^2\rangle C^2 + \langle \tilde{v}_I^2\rangle S^2 \\ -[\langle \tilde{u}_R\tilde{v}_I\rangle + \langle \tilde{u}_I\tilde{v}_R\rangle]CS & -2\langle \tilde{v}_R\tilde{v}_I\rangle CS \end{pmatrix}. \quad (18)$$

Here we used the shorthand notations

$$C \equiv \cos(\omega t), \, S \equiv \sin(\omega t). \quad (19)$$

Mean field and covariance tensor are all that is necessary to present the nonstationary bivariate Gaussian PDF $P(E_x, E_y, t)$ as

$$P(E_x, E_y, t) = \frac{1}{2\pi\sqrt{\langle \tilde{E}_x^2(t)\rangle\langle \tilde{E}_y^2(t)\rangle[1 - \rho^2(t)]}}$$
$$\times \exp\left[-\frac{(E_x - \langle E_x(t)\rangle)^2}{2[1 - \rho^2(t)]\langle \tilde{E}_x^2(t)\rangle} - \frac{(E_y - \langle E_y(t)\rangle)^2}{2[1 - \rho^2(t)]\langle \tilde{E}_y^2(t)\rangle}\right. \quad (20)$$
$$\left. - \frac{\rho(t)(E_x - \langle E_x(t)\rangle)(E_y - \langle E_y(t)\rangle)}{\sqrt{\langle \tilde{E}_x^2(t)\rangle\langle \tilde{E}_y^2(t)\rangle}[1 - \rho^2(t)]}\right],$$

where the correlation coefficient of the fluctuating field components $\tilde{E}_x$ and $\tilde{E}_y$ is

$$\rho(t) = \frac{\langle \tilde{E}_x(t)\tilde{E}_y(t)\rangle}{\sqrt{\langle \tilde{E}_x^2(t)\rangle\langle \tilde{E}_y^2(t)\rangle}}$$
$$= \frac{\langle \tilde{u}_R\tilde{v}_R\rangle C^2 - [\langle \tilde{u}_R\tilde{v}_I\rangle + \langle \tilde{u}_I\tilde{v}_R\rangle]CS + \langle \tilde{u}_I\tilde{v}_I\rangle S^2}{\sqrt{[\langle \tilde{u}_R^2\rangle C^2 - 2\langle \tilde{u}_R\tilde{u}_I\rangle CS + \langle \tilde{u}_I^2\rangle S^2][\langle \tilde{v}_R^2\rangle C^2 - 2\langle \tilde{v}_R\tilde{v}_I\rangle CS + \langle \tilde{v}_I^2\rangle S^2]}}. \quad (21)$$

Nonstationary two-dimensional PDF $P(E_x, E_y, t)$, Eq. (20) is a periodic function of time with period $t_O$. The mean electric vector $\langle \mathbf{E}(t) \rangle$ traces out a mean polarization ellipse with period $t_O$ while, according to Eq. (18), the scattering ellipse changes its shape, size and orientation with period $t_O/2$. Fig. 2 shows two examples of $P(E_x, E_y, t)$ at five $t_O/5$ -spaced time moments starting with $t=0$. For both examples the mean field is

$$\begin{pmatrix} \langle u_R \rangle \\ \langle u_I \rangle \\ \langle v_R \rangle \\ \langle v_I \rangle \end{pmatrix} = \begin{pmatrix} 10 \\ 2 \\ 0 \\ 20 \end{pmatrix}, \quad (22)$$

and the mean field ellipse is the same for both panels. At the left panel the scattering ellipse keeps the constant size and shape while rotating 360°. Statistical parameters for the left panel of Fig. 2 are

$$\mathbf{C} = \begin{pmatrix} 10 & 1 & -1 & 4 \\ 1 & 3.3 & 2.7 & 1 \\ -1 & 2.7 & 2 & -1 \\ 2 & 1 & -1 & 10 \end{pmatrix}, \mathbf{S} = \begin{pmatrix} 530.6 \\ -296 \\ 80 \\ -402.6 \end{pmatrix}, \mathbf{S}_{POL} = \begin{pmatrix} 506.1 \\ -296 \\ 80 \\ -402.6 \end{pmatrix},$$

$$\mathbf{S}_{DEP} = \begin{pmatrix} 24.5 \\ 0 \\ 0 \\ 0 \end{pmatrix}, \mathbf{S}_M = \begin{pmatrix} 504 \\ -296 \\ 80 \\ -400 \end{pmatrix}, \mathbf{S}_F = \begin{pmatrix} 26.6 \\ 0 \\ 0 \\ -2.6 \end{pmatrix}, \quad (23)$$

and the polarization degree is 0.954. Statistical parameters for the right panel of Fig. 2 are

$$\mathbf{C} = \begin{pmatrix} 10 & 0 & 0 & 0 \\ 0 & 0.3 & 0 & 0 \\ 0 & 0 & 0.3 & 0 \\ 0 & 0 & 0 & 2 \end{pmatrix}, \mathbf{S} = \begin{pmatrix} 516.6 \\ -288 \\ 80 \\ -400 \end{pmatrix}, \mathbf{S}_{POL} = \begin{pmatrix} 499.3 \\ -288 \\ 80 \\ -400 \end{pmatrix},$$

$$\mathbf{S}_{DEP} = \begin{pmatrix} 17.3 \\ 0 \\ 0 \\ 0 \end{pmatrix}, \mathbf{S}_M = \begin{pmatrix} 504 \\ -296 \\ 80 \\ -400 \end{pmatrix}, \mathbf{S}_F = \begin{pmatrix} 12.6 \\ 8 \\ 0 \\ 0 \end{pmatrix}, \quad (24)$$

and the polarization degree is 0.967. For this example the principal axes of the scattering ellipse maintain their orientation while the major and minor axes change according to Eq. (18). Note, that following the convention described earlier, $\langle v_R \rangle = 0$, applies to this and all examples that follow. For Fig. 2 this simply implies that at the moment $t=0$ the center of the scattering ellipse is located at the $x$ – axis.

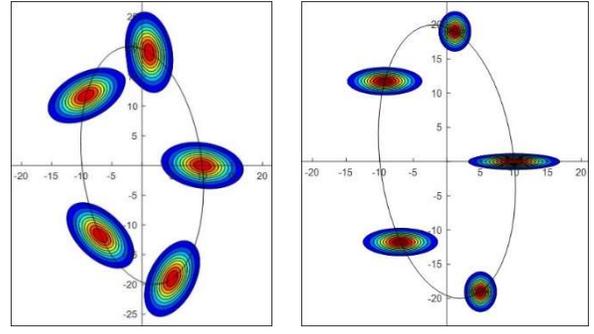

Fig. 2. Examples of the of the nonstationary PDF changes over the oscillation period. Left panel: rotating scattering ellipse with constant principal axes. Right panel: principal axes maintain direction but change the magnitude.

Figure 3 shows two contrasting examples of the PA-PDF calculated by numerical time averaging of the nonstationary Gaussian PDF, Eq. (20), over the oscillation period $t_O$, as prescribed by Eq. (8).

The left panel depicts the PA-PDF for the almost fully elliptical polarized field with very small isotropic fluctuating component. Statistical parameters for the left panel of Fig. 3 are

$$\begin{pmatrix} \langle u_R \rangle \\ \langle u_I \rangle \\ \langle v_R \rangle \\ \langle v_I \rangle \end{pmatrix} = \begin{pmatrix} -1 \\ 0 \\ 0 \\ 2 \end{pmatrix}, \mathbf{C} = \begin{pmatrix} 0.001 & 0 & 0 & 0 \\ 0 & 0.001 & 0 & 0 \\ 0 & 0 & 0.001 & 0 \\ 0 & 0 & 0 & 0.001 \end{pmatrix},$$

$$\mathbf{S} = \begin{pmatrix} 5.004 \\ -3 \\ 0 \\ 4 \end{pmatrix}, \mathbf{S}_{POL} = \mathbf{S}_M = \begin{pmatrix} 5 \\ -3 \\ 0 \\ 4 \end{pmatrix}, \mathbf{S}_{DEP} = \mathbf{S}_F = \begin{pmatrix} 0.004 \\ 0 \\ 0 \\ 0 \end{pmatrix}, \quad (25)$$

and the degree of polarization in this case is 0.9992. Here the field vector resides mostly on the ellipse corresponding to the elliptically polarized mean field as should be expected. Somewhat unexpected is a higher probability to register the field vector in the $y$-polarized state than in the $x$-polarized state. This rather counterintuitive result can be readily explained by calculating the angular velocity of the elliptically polarized field vector and discovering that it is not constant, and is lowest at the larger semi axle position. Note also that the polarized part of the Stokes vector in this case is the same as the Stokes vector of the mean field and that the depolarized part of the Stokes vector is the same as the Stokes vector for the fluctuating field component.

The right panel of the Figure 3 shows the opposite case of the fully random (diffuse) field. Statistical parameters for the right panel of Fig. 3 are

$$\begin{pmatrix} \langle u_R \rangle \\ \langle u_I \rangle \\ \langle v_R \rangle \\ \langle v_I \rangle \end{pmatrix} = \begin{pmatrix} 0 \\ 0 \\ 0 \\ 0 \end{pmatrix}, \mathbf{C} = \begin{pmatrix} 1 & 0 & 0 & 0 \\ 0 & 0.001 & 0 & 0 \\ 0 & 0 & 0.001 & 0 \\ 0 & 0 & 0 & 1 \end{pmatrix},$$

$$\mathbf{S} = \mathbf{S}_{DEP} = \mathbf{S}_F = \begin{pmatrix} 2.002 \\ 0 \\ 0 \\ 0 \end{pmatrix}, \mathbf{S}_{POL} = \mathbf{S}_M = \begin{pmatrix} 0 \\ 0 \\ 0 \\ 0 \end{pmatrix}. \quad (26)$$

The mean field is absent for this case, and the Stokes vectors of the depolarized and fluctuating components are equal to the field Stokes vector while the polarized and mean parts of the Stokes vector are zero as is the degree of polarization. For this unpolarized wave one would expect to have a bivariate Gaussian distribution of the field, but the right panel of Fig. 3 shows that this is not the case. Notwithstanding the very simple structure of the Stokes vector, the PA-PDF shows a sharp peak and four ridges that are related to the disparity of the diagonal components of the covariance matrix **C**.

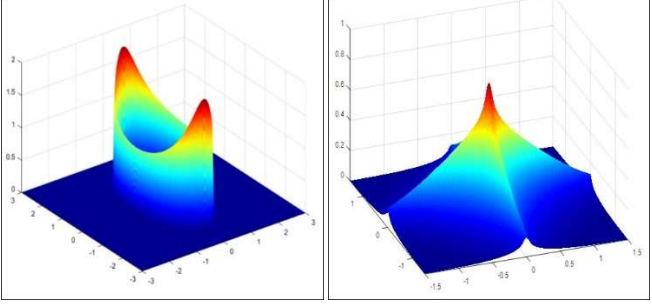

Fig. 3. Examples of PA-PDFs. Left panel: almost perfectly elliptically-polarized wave. Right panel: depolarized wave with non-Gaussian PA-PDF.

## 4. Diversity of polarized waves with fixed Stokes vector

As was mentioned earlier, only four combinations of the 14 statistical parameters of the nonstationary Gaussian enter Eq. (12) for the Stokes vector. This indicates that the Stokes vector provides incomplete characterization of the partially-polarized waves statistics and opens up the possibility of altering the PA-PDF without changing the Stokes vector. Three fairly straightforward ways to do this are discussed here. However, a full examination of the nine-dimensional $\mathbf{S} = const$ manifold in the 13-dimensional space of the PA-PDF parameters is outside the scope of this paper.

### A. Mean values split

Examination of Eq. (12) or Eq. (13) suggests that the products of the mean values of the quadratic components of the field always enter the components of the Stokes vector as a sum with the corresponding covarinces, e. g.

$$\langle u_R \rangle^2 + \langle \widetilde{u}_R^2 \rangle, \ \langle u_R \rangle \langle v_I \rangle + \langle \widetilde{u}_R \widetilde{v}_I \rangle, \ \dots \quad (27)$$

This allows a simultaneous change to the mean values and the covariances while maintaining the fixed value of the Stokes vector. Of course, these manipulations should preserve the SPD property of the covariance matrix. Without getting into the details of the pertinent limitations, we show an example of such a transformation in Fig. 4. For both panels of Fig. 4, the Stokes vector and its classical polarized and depolarized components are

$$\mathbf{S} = \begin{pmatrix} 30.7 \\ 20.5 \\ 0 \\ -20 \end{pmatrix}, \ \mathbf{S}_{POL} = \begin{pmatrix} 28.6 \\ 20.5 \\ 0 \\ -20 \end{pmatrix}, \ \mathbf{S}_{DEP} = \begin{pmatrix} 2.1 \\ 0 \\ 0 \\ 0 \end{pmatrix}, \quad (28)$$

and the polarization degree is 0.933. The rest of the statistical parameters for the left panel of Fig. 4 are

$$\begin{pmatrix} \langle u_R \rangle \\ \langle u_I \rangle \\ \langle v_R \rangle \\ \langle v_I \rangle \end{pmatrix} = \begin{pmatrix} 5 \\ 0 \\ 0 \\ 2 \end{pmatrix}, \ \mathbf{C} = \begin{pmatrix} 0.5 & 0 & 0 & 0 \\ 0 & 0.1 & 0 & 0 \\ 0 & 0 & 0.1 & 0 \\ 0 & 0 & 0 & 1 \end{pmatrix},$$

$$\mathbf{S}_M = \begin{pmatrix} 29 \\ 21 \\ 0 \\ -20 \end{pmatrix}, \ \mathbf{S}_F = \begin{pmatrix} 1.7 \\ -0.5 \\ 0 \\ 8 \end{pmatrix}, \quad (29)$$

and for the right panel of Fig. 4 statistical parameters are

$$\begin{pmatrix} \langle u_R \rangle \\ \langle u_I \rangle \\ \langle v_R \rangle \\ \langle v_I \rangle \end{pmatrix} = \begin{pmatrix} 4.55 \\ 0 \\ 0 \\ 2.22 \end{pmatrix}, \ \mathbf{C} = \begin{pmatrix} 5.25 & 0 & 0 & 0 \\ 0 & 0.1 & 0 & 0 \\ 0 & 0 & 0.1 & 0 \\ 0 & 0 & 0 & 0.0617 \end{pmatrix},$$

$$\mathbf{S}_M = \begin{pmatrix} 25.2 \\ 15.3 \\ 0 \\ -20 \end{pmatrix}, \ \mathbf{S}_F = \begin{pmatrix} 5.5 \\ 5.2 \\ 0 \\ 0 \end{pmatrix}. \quad (30)$$

Changes to the mean field and covariance matrix preserve the Stokes vector and its polarized and depolarized components, but the mean and fluctuating components, of the Stokes vector, Eq. (15), vary.

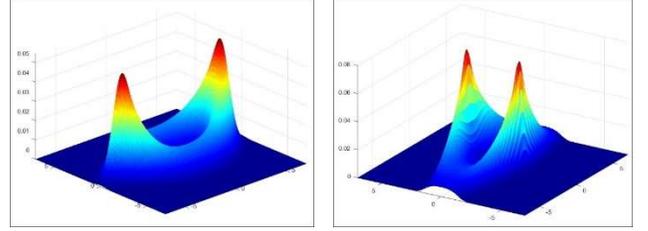

Fig. 4. Examples of PA-PDFs with different mean fields but identical Stokes vectors.

The dramatic change of the PA-PDF is obvious. In particular, the most probable state of the field changes from the *y*-polarized to *x*-polarized while the Stokes vector remains unchanged.

### B. Quadrature components split

It follows from Eq. (12) that the coherence tensor, and, hence the Stokes vector remain unchanged when $\langle u_R^2 \rangle + \langle u_I^2 \rangle$, $\langle u_R v_R \rangle + \langle u_I v_I \rangle$, and $\langle u_R v_I \rangle - \langle u_I v_R \rangle$ remain constant, even when the individual components of the mean field and the covariance matrix **C** vary. There are many different ways to do this, and without getting into the details of the all possible transformations, we present two examples here. For all charts in the Fig. 5 the mean field, Stokes vector and the Stokes vector components are

$$\begin{pmatrix}\langle u_R\rangle\\\langle u_I\rangle\\\langle v_R\rangle\\\langle v_I\rangle\end{pmatrix}=\begin{pmatrix}2\\0\\0\\5\end{pmatrix},\ \mathbf{S}=\begin{pmatrix}29.7\\-21.3\\0\\-20.3\end{pmatrix},\ \mathbf{S}_{POL}=\begin{pmatrix}29.4\\-21.3\\0\\-20.3\end{pmatrix},$$

$$\mathbf{S}_{DEP}=\begin{pmatrix}0.3\\0\\0\\0\end{pmatrix},\ \mathbf{S}_M=\begin{pmatrix}29\\-21\\0\\-20\end{pmatrix},\ \mathbf{S}_F=\begin{pmatrix}0.7\\-0.3\\0\\-0.3\end{pmatrix},$$

(31)

and the polarization degree is 0.990.

For the left panel of Fig. 5 the covariance matrix of the fluctuating components is

$$\mathbf{C}=\begin{pmatrix}0.21 & 0 & 0 & 0.16\\ 0 & 0.01 & 0 & 0\\ 0 & 0 & 0.01 & 0\\ 0.16 & 0 & 0 & 0.51\end{pmatrix},$$

(32)

and PA-PDF has four modes of equal height at ±80° and ±100°.

For the right panel of Fig. 5 the covariance matrix of the fluctuating component is

$$\mathbf{C}=\begin{pmatrix}0.01 & 0 & 0 & 0\\ 0 & 0.21 & -0.16 & 0\\ 0 & -0.16 & 0.51 & 0\\ 0 & 0 & 0 & 0.01\end{pmatrix},$$

(33)

and PA-PDF has four modes of unequal height at 0°, 180° and ±90°. Note that the three combination of the covariances $\langle\tilde{u}_R^2\rangle+\langle\tilde{u}_I^2\rangle$, $\langle\tilde{u}_R\tilde{v}_R\rangle-\langle\tilde{u}_I\tilde{v}_I\rangle$, and $\langle\tilde{u}_R\tilde{v}_I\rangle+\langle\tilde{u}_I\tilde{v}_R\rangle$ are the same for both Eq. (32) and Eq. (33).

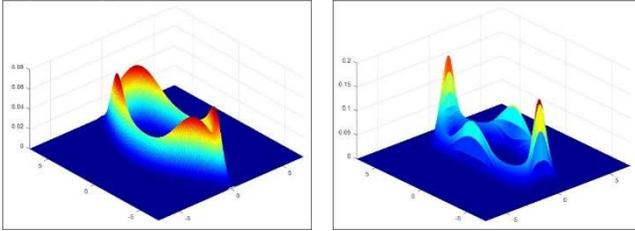

Fig. 5. Examples of PA-PDFs with identical mean fields and Stokes vectors but varying quadrature components.

### C. Covariance of the in-phase and quadrature components

It is clear from the Eq. (12) that covariances $\langle\tilde{u}_R\tilde{u}_I\rangle$ and $\langle\tilde{v}_R\tilde{v}_I\rangle$ do not affect the Stokes vector. Therefore it is possible to change the shape of the PA-PDF while maintaining the fixed Stokes vector and degree of polarization by altering these covariances. It is still necessary to maintain the SPD property of the covariance matrix **C**, which puts certain limitations on the choice of $\langle\tilde{u}_R\tilde{u}_I\rangle$ and $\langle\tilde{v}_R\tilde{v}_I\rangle$. For all charts in the Fig. 6 the mean field, the Stokes vector and its components are

$$\begin{pmatrix}\langle u_R\rangle\\\langle u_I\rangle\\\langle v_R\rangle\\\langle v_I\rangle\end{pmatrix}=\begin{pmatrix}5\\0\\0\\5\end{pmatrix},\ \mathbf{S}=\begin{pmatrix}62\\0\\0\\-50\end{pmatrix},$$

$$\mathbf{S}_{POL}=\mathbf{S}_M=\begin{pmatrix}50\\0\\0\\-50\end{pmatrix},\ \mathbf{S}_{DEP}=\mathbf{S}_F=\begin{pmatrix}12\\0\\0\\0\end{pmatrix},$$

(34)

and the degree of polarization is 0.81. Based on the Eq. (34), the mean field is circular polarized, and there is a depolarized fluctuating component.

For the left panel of Fig. 6 the covariance matrix of the fluctuating component is

$$\mathbf{C}=\begin{pmatrix}3 & 0 & 0 & 0\\ 0 & 3 & 0 & 0\\ 0 & 0 & 3 & 0\\ 0 & 0 & 0 & 3\end{pmatrix},$$

(35)

and PA-PDF is circular-symmetric and has a toroidal shape. This is probably the simplest example of the field satisfying Eq. (34) where the fluctuating component has an isotropic and stationary covariance tensor, Eq. (18).

For the central panel of Fig. 6, the covariance matrix is:

$$\mathbf{C}=\begin{pmatrix}3 & -2.97 & 0 & 0\\ -2.97 & 3 & 0 & 0\\ 0 & 0 & 3 & 2.7\\ 0 & 0 & 2.7 & 3\end{pmatrix},$$

(36)

and the covariance tensor is nonstationary which causes the scattering ellipse to change the major and minor axes, while maintaining the axes direction. The resulting PA-PDF has an unusual, four-mode diamond shape.

For the right panel in Fig. 6 the covariance matrix is:

$$\mathbf{C}=\begin{pmatrix}3 & 2.85 & 0 & 0\\ 2.85 & 3 & 0 & 0\\ 0 & 0 & 3 & 2.85\\ 0 & 0 & 2.85 & 3\end{pmatrix},$$

(37)

and the covariance tensor, Eq. (18), again is nonstationary causing the scattering ellipse to change its size, while maintaining the circular symmetry (see Fig. 2 for reference). The resulting PA-PDF has two well-defined modes, and is very dissimilar to the both of the previous examples.

The examples shown in Figs. 4 - 6 clearly demonstrate that the Stokes vector does not provide the full information regarding the field statistics. Three approaches presented here by no means exhaust all the possible variations of statistics of the partially-polarized wave with fixed Stokes vector.

### 5. Discussion

Analytical development illustrated by several examples in the previous section demonstrated that the Stokes vector provides only partial information about the state of the partially polarized narrow-band electromagnetic field. This is in stark contrast with Van de Hulst's [6, p. 43] Principle of Optical Equivalence: "*It is impossible by means of any instruments to distinguish between various incoherent sums of simple waves that may together form a beam with the same Stokes parameters.*"

The explanation for this apparent disagreement is very simple: Stokes and his countless followers considered a simple incoherent measurement system, typically including phase retarder, polarizer and a photodetector [3, 7]. The measurements of the 13 statistical parameters introduced here requires coherent (heterodyne) detection.

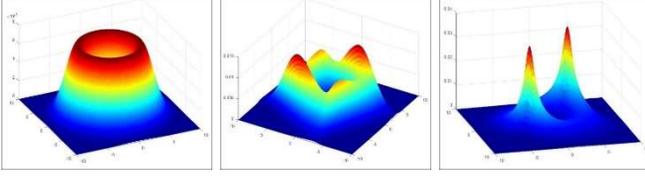

Fig. 6. Examples of PA-PDFs with the identical mean fields and Stokes vectors, but varying covariances $\langle \tilde{u}_R \tilde{u}_I \rangle$ and $\langle \tilde{v}_R \tilde{v}_I \rangle$.

Consider two-channel heterodyne detection system where in the *x*-channel the partially polarized narrow-band field is mixed with the *x* and *y*-polarized local oscillator (LO) fields

$$\mathbf{E}_X^{(LO)}(t) = \begin{pmatrix} U \\ 0 \end{pmatrix} \cos[(\omega + \Omega)t + \varphi],$$

$$\mathbf{E}_Y^{(LO)}(t) = \begin{pmatrix} 0 \\ V \end{pmatrix} \cos[(\omega + \Omega)t + \varphi]. \quad (38)$$

Here $\Omega \ll \omega$ is an intermediate (radio) frequency, $\Omega t_C \gg 1$, and we assume that the LO phase $\varphi$ is unknown but remains stable during the measurement time $T \gg t_C$. The detector photocurrent for the *x*–channel is proportional to the square of the sum of the signal and LO fields

$$I^{(D)}(t) = R[U\cos[(\omega+\Omega)t+\varphi] + u_R(t)\cos(\omega t) - u_I(t)\sin(\omega t)]^2 \\ + R[v_R(t)\cos(\omega t) - v_I(t)\sin(\omega t)]^2 \quad (39)$$

where R is detector responsivity. The detector signal is comprised of a DC component proportional to $U^2$, slowly changing at $t_C$ time scale component proportional to $u_{R,I}^2, v_{R,I}^2$, and slowly changing IF component $I^{(\Omega)}(t)$ at the frequency $\Omega$ for the *x*-channel, and similar component $J^{(\Omega)}(t)$ for the *y*-channel which are of interest here:

$$I^{(\Omega)}(t) = RU[u_R(t)\cos\varphi + u_I(t)\sin\varphi]\cos(\Omega t) \\ - RU[u_I(t)\cos\varphi - u_R(t)\sin\varphi]\sin(\Omega t), \\ J^{(\Omega)}(t) = RV[v_R(t)\cos\varphi + v_I(t)\sin\varphi]\cos(\Omega t) \\ - RV[v_I(t)\cos\varphi - v_R(t)\sin\varphi]\sin(\Omega t). \quad (40)$$

The fast oscillating component at frequency $2\omega_O$ is not registered by detector, but we assume that the IF frequency signals $I^{(\Omega)}(t)$ and $J^{(\Omega)}(t)$ can be sampled properly. Analytical signals $I^{(AS)}(t)$ and $J^{(AS)}(t)$ with carrier frequency $\Omega$ corresponding to the real signals $I^{(\Omega)}(t)$ and $J^{(\Omega)}(t)$ can be recovered using the Hilbert transformation of the time series for $I^{(\Omega)}(t)$ and $J^{(\Omega)}(t)$ as

$$I^{(AS)}(t) = RU\{[u_R(t)\cos\varphi + u_I(t)\sin\varphi] \\ + i[u_I(t)\cos\varphi - u_R(t)\sin\varphi]\}\exp(i\Omega t), \\ J^{(AS)}(t) = RV\{[v_R(t)\cos\varphi + v_I(t)\sin\varphi] \\ + i[v_I(t)\cos\varphi - v_R(t)\sin\varphi]\}\exp(i\Omega t). \quad (41)$$

Note that the unknown LO phase $\varphi$ is assumed to be the same for both channels. Down converting these analytical signals to DC produces two random complex signals with $t_C$ time scale

$$I^{(C)}(t) = I_R(t) + iI_I(t) = RU[u_R(t)\cos\varphi + u_I(t)\sin\varphi] \\ + iRU[u_I(t)\cos\varphi - u_R(t)\sin\varphi], \\ J^{(C)}(t) = J_R(t) + iJ_I(t) = RV[v_R(t)\cos\varphi + v_I(t)\sin\varphi] \\ + iRV[v_I(t)\cos\varphi - v_R(t)\sin\varphi]. \quad (42)$$

The time series for $I_R(t), I_I(t), J_R(t)$, and $J_I(t)$ can be used to estimate the statistics of these four real signals including the mean values and covariance matrix.

Formally, Eq. (42) can be solved for $u_R(t), u_I(t), v_R(t)$, and $v_I(t)$ as follows

$$u_R(t) = \frac{1}{RU}I_R(t)\cos\varphi - \frac{1}{RU}I_I(t)\sin\varphi, \\ u_I(t) = \frac{1}{RU}I_I(t)\cos\varphi + \frac{1}{RU}I_R(t)\sin\varphi, \\ v_R(t) = \frac{1}{RV}J_R(t)\cos\varphi - \frac{1}{RV}J_I(t)\sin\varphi, \\ v_I(t) = \frac{1}{RV}J_I(t)\cos\varphi + \frac{1}{RV}J_R(t)\sin\varphi. \quad (43)$$

However, the field components $u_R(t), u_I(t), v_R(t)$, and $v_I(t)$ are still not accessible since the LO phase $\varphi$ is unknown. The unknown phase can be eliminated if an additional assumption is imposed on the statistics of the field. This is equivalent to the uncertainty of the measurements timing discussed earlier that led to the PA-PDF concept and the reduction of the number of statistical parameters from 14 to 13. Using the same, arbitrary, prescription $\langle v_R \rangle = 0$, we set

$$\varphi = \arctan\frac{\langle J_R \rangle}{\langle J_I \rangle} \quad (44)$$

This determines the sought quadrature components of the polarized field $u_R(t), u_I(t), v_R(t)$, and $v_I(t)$ in terms of the measured quadrature components of the IF photocurrent as

$$u_R(t) = \frac{1}{RU} \frac{\langle J_I \rangle I_R(t) - \langle J_R \rangle I_I(t)}{\sqrt{\langle J_R \rangle^2 + \langle J_I \rangle^2}},$$

$$u_I(t) = \frac{1}{RU} \frac{\langle J_I \rangle I_I(t) + \langle J_R \rangle I_R(t)}{\sqrt{\langle J_R \rangle^2 + \langle J_I \rangle^2}},$$

$$v_R(t) = \frac{1}{RV} \frac{\langle J_I \rangle J_R(t) - \langle J_R \rangle J_I(t)}{\sqrt{\langle J_R \rangle^2 + \langle J_I \rangle^2}},$$

$$v_I(t) = \frac{1}{RV} \frac{\langle J_I \rangle J_I(t) + \langle J_R \rangle J_R(t)}{\sqrt{\langle J_R \rangle^2 + \langle J_I \rangle^2}}.$$

(45)

The time series for the quadrature components of the polarized field $u_R(t), u_I(t), v_R(t)$ and $v_I(t)$ can now be used to calculate the mean values and the covariance matrix **C**, Eq. (11).

The PA-PDF can be estimated from the samples of the scaled IF real signals $I^{(\Omega)}(t)/RU$ and $J^{(\Omega)}(t)/RV$. The phase $\varphi$ is irrelevant here since it corresponds to an unknown, but fixed shift of the clock zero.

## 6. Conclusion

Considering the electrical field of a partially polarized wave as a two-dimensional real oscillating random vector, we introduced the PA-PDF that encompasses the complete statistics of the field at a single point.

For the simple case of a nonstationary oscillating Gaussian probability distribution of the field, we find that field statistics at a single point are completely described by 13 parameters. This is in contrast to the four parameters provided by the conventional Stokes vector or coherence tensor descriptions.

Using several examples, we illustrated that the partially polarized field state as described by PA-PDF can vary significantly when the Stokes vector remains fixed.

PA-PDF and nine statistical parameters supplementary to the Stokes vector can be measured by a two-channel heterodyne detectors and require a stable local oscillator.

Our results suggest that it is possible to extract a lot more information from the partially polarized wave than is currently though of based on the Stokes vector description.

We believe that this new development can be important for a variety of remote sensing applications where medium or surface parameters are derived from the statistics of the scattered electromagnetic waves or where partially polarized waves are used as a probe.

For example, this new development can be used for remote sensing of particular matter, such as clouds [8] and aerosols, as well as for the radar and optical signatures of the randomly rotating targets. Other applications are optical and radar surveillance of the sea surface, surface terrain and vegetation [9], as well as imaging and sensing of biological tissues [10].

This presented theory is not limited to the optical range, but can also be applied to the radar polarimetry and polarimetric SAR interferometry [9, 11].

Thirteen measurable degrees of freedom of the partially polarized wave, even when restricted by the SPD constraint, make this type of wave an attractive choice for the optical communication, but this topic is out of the scope of this paper.

This work addresses the statistics of the partially polarized field at a single point. The issues related to the spatial coherence of the field which are essential for the development of the propagation model are not discussed. However, it becomes clear that the conventional, coherence tensor-based formulation, of the partially polarized waves' propagation [12] cannot fully describe their statistics.

In this work Cartesian basis is used for the polarized waves, representation. It would be interesting to use the right/left circular polarized waves as the basis.

**Acknowledgment**. Author thanks Deanna F. for editing this manuscript.